\documentclass[journal]{IEEEtran}
\ifCLASSINFOpdf
\else
\fi

\usepackage{cite}
\usepackage{graphicx}
\usepackage{setspace}
\usepackage{subfigure}
\usepackage{amsmath}
\usepackage{caption}
\begin{document}
%

\title{A $3.8~ps$ RMS time synchronization implemented in a $20~nm$ FPGA}
%
%
%

\author{Hong-Bo Xie, Yang Li, Qi Shen, Sheng-Kai Liao, Cheng-Zhi Peng
\thanks{This work was supported by the China Postdoctoral Science Foundation under Grant No. 2016M600481,
Natural Science Foundation of Shanghai under Grants 18ZR1443600, and 18ZR1443700, Anhui Provincial Natural Science Foundation under Grant 1808085QF180, and National Natural Science Foundation of China under Grant 11705191. }

\thanks{Authors are with the Hefei National Laboratory for Physical Sciences at the Microscale and Department of Modern Physics, University of Science and Technology of China, Hefei 230026, China, and Chinese Academy of Sciences (CAS) Center for Excellence and Synergetic Innovation Center in Quantum Information and Quantum Physics, University of Science and Technology of China, Shanghai 201315, China (e-mail: $xiehongb@mail.ustc.edu.cn$, $liyang9@ustc.edu.cn$, $shenqi@ustc.edu.cn$, $skliao@ustc.edu.cn$).}

}

%


\maketitle

\begin{abstract}
A $3.8~ps$ root mean square (RMS) time synchronization implemented in a $20~nm$ fabrication process ultrascale kintex Field Programmable Gate Array (FPGA) is presented. The multichannel high-speed serial transceivers ($e.g.$, GTH) play a key role in a wide range of applications, such as the optical source for quantum key distribution systems. However, owing to the independent clock dividers existed in each transceiver, the random skew would appear among the multiple channels every time the system powers up or resets. A self-phase alignment method provided by Xilinx Corporation could reach a precision with $22~ps$ RMS and $100~ps$ maximum variation, which is far from meeting the demand of applications with rate up to $2.5~Gbps$. To implement a high-precision intrachannel time synchronization, a protocol combined of a high-precision time-to-digital converter (TDC) and a tunable phase interpolator (PI) is presented. The TDC based on the $carry8$ primitive is applied to measure the intrachannel skew with $40.7~ps$ bin size. The embedded tunable PI in each GTH channel has a theoretical step size of $3.125~ps$. By tuning the PI in the minimal step size, the final intrachannel time synchronization reaches a $3.8~ps$ RMS precision and maximal variation $20~ps$, much better than the self-phase alignment method. Besides, a desirable time offset of every channel can be implemented with a closed-loop control.
\end{abstract}


\IEEEpeerreviewmaketitle

\section{Introduction}
The success of Satellite-bound Quantum Key Distribution (QKD) paves a way to achieve a global QKD network \cite{Sate:Bound:2017}. Currently, the popular QKD optical source schemes mainly include two methods: one is the multi-lasers design \cite{Sour:multilaser:2014}. The other one is the single-laser method working with phase modulation and intensity modulation \cite{Sour:singlelaser:2010}. Both methods share a common demand that multichannel electric excitation signals with a relation of deterministic phase. For a low-speed ($\textless100~MHz$) application, a common clock is used to guarantee the phase synchronization and system jitter is mainly depends on the jitter of clock which is normally small to neglect. In high-speed ($\textgreater1~Gbps$) area, the parallel signals with a common clock have a large crosstalk and thus a serial transmission is adopted. The high-speed serial transceiver ($e.g.$ GTH) embedded in FPGA provides a simple and feasible solution for up to tens of gigahertz (GHz) applications \cite{Xilinx:docu:2016}. However, owing to the independent clock network in each GTH channel, it would result to a random skew among the multiple channels every time the system powers up or resets, which is a forbidden in some scenarios, especially for the high-speed QKD. In the case that the same external reference clock and same data rates of each channel, the random skew is still observed among the GTH channels when system powers on or resets. The provided method of channel alignment by GTH itself just reach a precision $22~ps$ and a maximum variation $100~ps$, which is far from the demand of applications that needs $400~ps$ electric pulse and $\textless40~ps$ intrachannel skew jitter. Fortunately, there is a PI in the clock path, and we can adjust it to vary intrachannel delay precisely. Another key factor is that obtaining the actual skew among multichannel GTHs. Interpolator-based TDC has been proven to be a significant way to realize a high-precision measurement.

Time-interval TDC implemented in FPGA is adopting a coarse counter that runs at the system clock and a fine measurement using time interpolation to yield a sub-clock period resolution. The mainly idea of time interpolator is that the hit signal propagates along a tapped delay line (TDL) and the status of TDL was recorded at the rising edge of system clock. The architecture that TDL consists of cascade of fast carry-chains was first used by Wu in 2003 \cite{TDC:Wujin:2003} and a $400~ps$ bin size was obtained in Altera ACEK FPGA. From then on, a large number of studies about TDC using carry-chains as TDL were published \cite{TDC:CFavi:2009}\cite{TDC:Wujin:2008}. The achievable time resolution has reach an order of $4.2~ps$ resolution in a single TDC in kintex ultraScale FPGA \cite{TDC:WangY:2016}, which is the highest resolution reported. Besides, the averaging method of multiple measurements and multi-chain measurements can further improve the resolution and precision. In 2015, Shen et al. implemented a $1.7~ps$ equivalent bin size and $4.2~ps$ RMS resolution in a $40~nm$ fabrication process Virtex-6 FPGA \cite{TDC:Shen:2015}. The TDL¡¯s architecture based on the carry chain has been proven to be an appropriate way to implement an FPGA-based TDC.

In this paper, we implement a high-precision $3.8~ps$ RMS time synchronization among the multichannel transceivers with the PI and TDC in a closed loop. Additional, an arbitrary time offsets among channels can be achieved.

\begin{figure*}[!htbp]
 \centering
 \includegraphics[width=17cm]{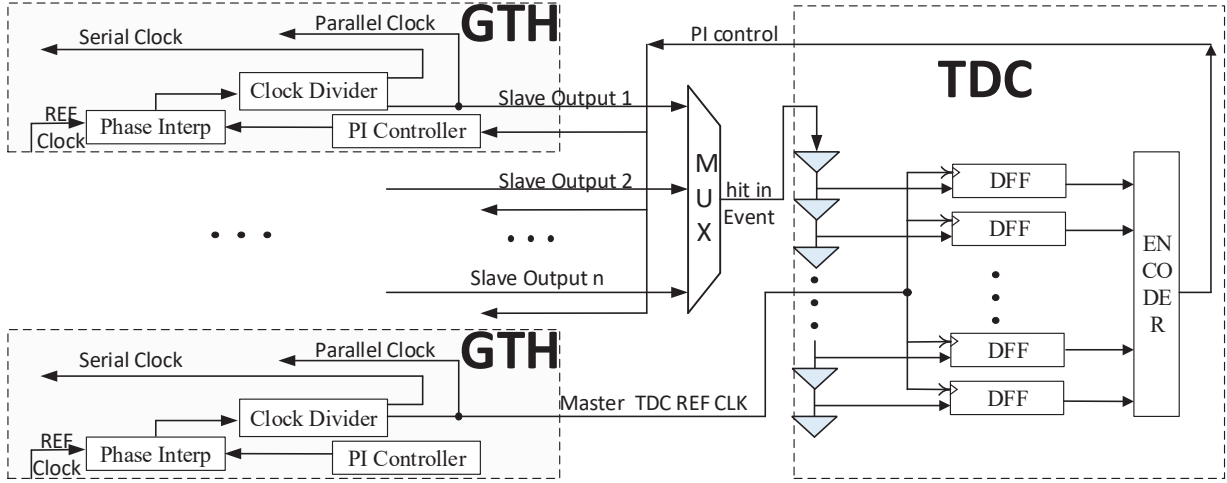}
 \caption{The structure of system: a center GTH channel is labelled as the master lane and its parallel clock is routed as the sampling clock of TDC. The parallel clock of other channels as slaves are combined into a multiplexer (MUX). The selected output clock is divided to generate a low-speed signal to act as the hit event of TDC. The readout temperature code of TDL is encoded to a bin code. The measured result decides the behavior of PI. }
\label{Fig:System:Structure}
\end{figure*}

The remaining part of paper is organized as follows. The methods of intrachannel synchronization is depicted in Section II. In Section III, the test result about the performance of system is described. Finally, the conclusion is given in Section IV.

\section{Time synchronization methods}
The block diagram of multichannel synchronization is shown in Fig. \ref{Fig:System:Structure}. It consists of a TDC and eight GTH channels. All GTH channels share a same external clock and then generate the respective serial clocks as well as parallel clocks. One center channel is labelled as the master channel and its parallel clock is routed to act as the sampling clock of TDC. The other channels is regarded as the slave channels. All slave channels¡¯ parallel clocks are combined into a mux and only one is selected to output at the same time. The selected output clock of muxer is divided to generate a periodic low-frequency signal which is assumed to have a certain phase relationship with selected clock. The hit signal is routed into the TDC and propagate along the TDL. The status of TDC is read out at the rising edge of sampling clock. The ideal readout data is a temperature code, but it always exists so-called bubbles effect because of the routed deviation of sampling clock. Thus, we need to utilize the proper encode algorithm to eliminate the effect and turn readout data into a binary code. The encoded result is the rising-edge skew between the selected slave channel and the master channel. We can preset a target skew in advance, continuously tune the PI of slave channel until measured skew equals to target value in permission scope. Each time the system powers up or resets, the same alignment process is conducted. Finally, the time synchronization between the master channel and the slave channel is accomplished. With the same way, we can implemented the time synchronization among all channels.

The intrachannel synchronization precision depends on the precision of PI and the resolution of TDC. It has shown that the PI have a minimal step size to an order of $3.125~ps$ \cite{Xilinx:docu:2016}, which is enough to satisfy the demand of applications. As mentioned above, the self-phase alignment also uses the PI, but the time synchronization is still poor enough, which attributes to the precision of time measurement. TDC is capable of achieving a high-resolution measurement.

\section{System tests}
\subsection{TDC performance}

\begin{figure}[!htbp]
\centering
\begin{minipage}{8.5cm}
\centering
\includegraphics[width=8.5cm]{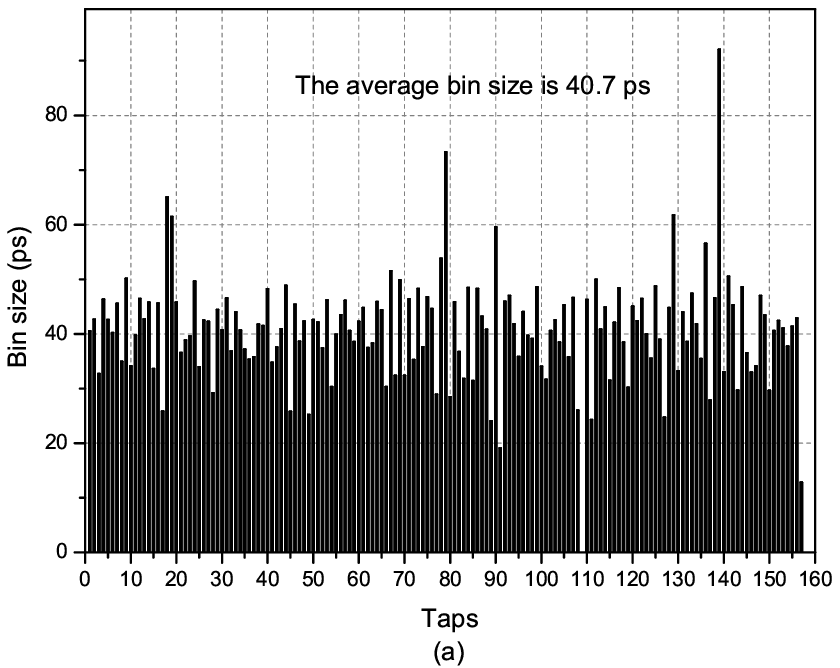}
\end{minipage}

\begin{minipage}{8.5cm}
\centering
\includegraphics[width=8.5cm]{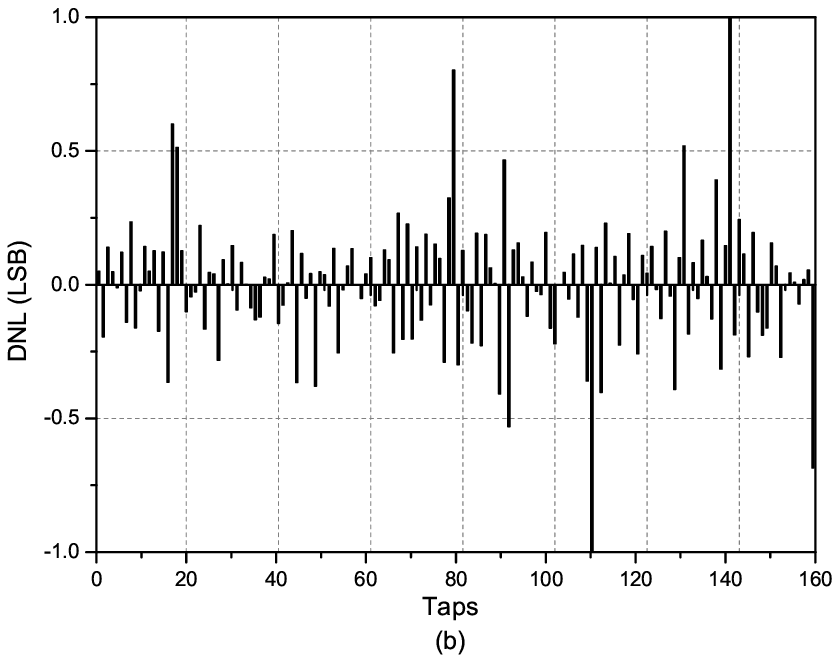}
\end{minipage}

\begin{minipage}{8.5cm}
\centering
\includegraphics[width=8.5cm]{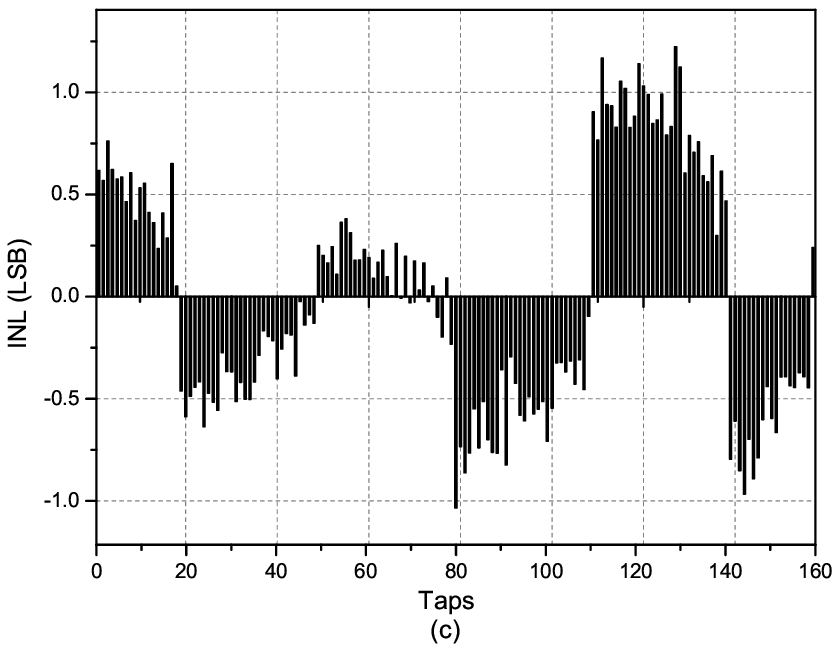}
\end{minipage}

\caption{Measured (a) bin size, (b) DNL, and (c) INL with Carry8 as minimal delay cell. }
\label{Fig:TDC:Charact}
\end{figure}

The fundamental carry unit is called $carry8$ in each slice of the Kintex Ultrascale FPGA. Cascading a number of $Carry8$ of different slices to implemented a TDL to guarantee the total delay longer than the system clock period. Test has indicated that the delay of 160 slices is just enough for the clock period of $6.4~ns$, and thus we construct a TDL using 160 slices locating in a column of FPGA. In theoretically, $carry8$ could be further divided into smaller delay unit. The number of taps differs with division of $carry8$. The smaller the bin size, the higher the resolution of TDC.
The characteristics of TDC, including bin width, DNL and INL were measured using the code density test. An independent oscillator was used to generate a hit event into TDC, and its frequency is non-correlated with the system clock of TDC. Owing to the hit signal appears in any position with the same probability, the relative fractions on each span reflect the physical bin width. Both INL and DNL can be extracted by further processing the sampling data of code density test. The Fig. \ref{Fig:TDC:Charact} shows the characteristics of TDC adopting $carry8$ as the minimal delay unit. The DNL in most bins are smaller than $¡À0.5$ LSB and the maximum is about 1.3 LSB. The fitting INL in most bins is smaller than $¡À1 LSB$, but appeals a periodic system error, which attributes to that the TDL spans banks in FPGA and the delay between banks is much bigger than the inter-bank delay cell. The INL would be better regardless of the effect of spanning banks. The average bin size is $40.7~ps$, which can be improved further through multiple measurement or multi-chain measurement.

\subsection{Time synchronization performance}

\begin{figure}[!htbp]
 \centering
 \includegraphics[width=8.5cm]{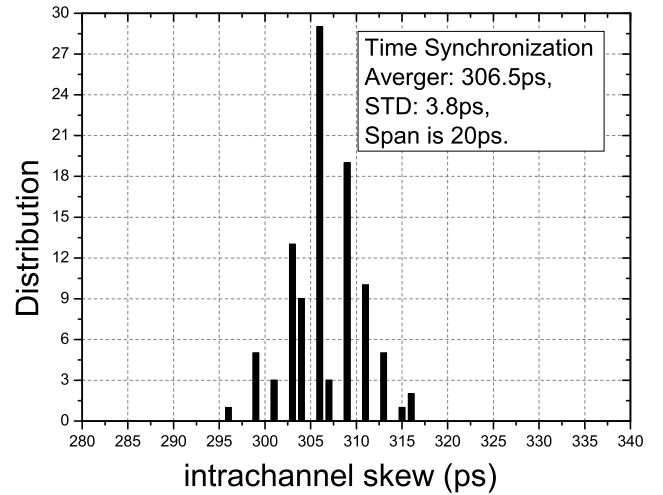}
 \caption{Intrachannel synchronization performance with TDC-based alignment. . }
\label{Fig:CHalign:Reso}
\end{figure}

A test for the performance of the channel synchronization is conducted between two GTH channels. The synchronization process are as depicted. Firstly, a target skew among channels is preset, and then we employ the TDC to measure the actual intrachannel skew. The difference between preset and target value is continuously decreased through moving the PI until the error in a tolerable range. Every time the system powers up or resets, a new synchronization process is started. We make a statistics about the delay between two channels after the locking of phase in a same point. It is amazing that a $3.8~ps$ RMS synchronization precision is achieved. The average is $306.5~ps$ and maximum variation is $¡À10~ps$, the result is shown in Fig. \ref{Fig:CHalign:Reso}. If there are more sample points, a perfect gauss distribution would be appealed. Meanwhile, another test has demonstrated the method of self-phase alignment can reach a precision with RMS $22~ps$ and the maximum variation $¡À50~ps$, worse than the method of TDC-based alignment.

\section{Conclusion}
We come out a novel approach to solve successfully the random skew between high-speed serial transceivers occurred when the system powers up or resets. More importantly, an extremely high-resolution (RMS $3.8~ps$) time synchronization is achievable among the transceivers running at rate of $2.5~Gbps$. Besides, we are capable of implementing arbitrary time difference with an approximate resolution among the multichannel GTHs. Such a high-precision time synchronization is very important in a wide range of applications, especially for high-speed quantum communication.

\ifCLASSOPTIONcaptionsoff
  \newpage
\fi

\end{document}